\def\year{2021}\relax
\documentclass[letterpaper]{article} 
\usepackage{aaai21}  
\usepackage{times}  
\usepackage{helvet} 
\usepackage{courier}  
\usepackage[hyphens]{url}  
\usepackage{graphicx} 
\usepackage{natbib}  
\usepackage{caption} 
\usepackage{subfigure}
\usepackage{booktabs}
\usepackage{amsfonts}
\usepackage{multirow}
\usepackage{amsmath}
\usepackage{algorithm}
\usepackage[noend]{algpseudocode}
\usepackage{tikz}

\newcommand{\eg}{e.\,g., }
\newcommand{\ie}{i.\,e., }

\definecolor{mygreen}{rgb}{0,0.6,0}
\definecolor{mymauve}{rgb}{0.58,0,0.82} 

\usetikzlibrary{arrows,shapes, decorations.pathmorphing,backgrounds,positioning}

\title{Transfer Graph Neural Networks for Pandemic Forecasting}
\author {

        George Panagopoulos,\textsuperscript{\rm 1}
        Giannis Nikolentzos, \textsuperscript{\rm 2}
        Michalis Vazirgiannis \textsuperscript{\rm 1} \\
}
\affiliations {
    \textsuperscript{\rm 1} \'Ecole Polytechnique, Palaiseau, France \\
    \textsuperscript{\rm 2} Athens University of Economics and Business, Athens, Greece \\
    george.panagopoulos@polytechnique.edu, \{nikolentzos, mvazirg\}@aueb.gr
}

\begin{document}

\maketitle

\begin{abstract}
The recent outbreak of COVID-19 has affected millions of individuals around the world and has posed a significant challenge to global healthcare. From the early days of the pandemic, it became clear that it is highly contagious and that human mobility contributes significantly to its spread.
In this paper, we study the impact of population movement on the spread of COVID-19, and we capitalize on recent advances in the field of representation learning on graphs to capture the underlying dynamics. Specifically, we create a graph where nodes correspond to a country's regions and the edge weights denote human mobility from one region to another. Then, we employ graph neural networks to predict the number of future cases, encoding the underlying diffusion patterns that govern the spread into our learning model. Furthermore, to account for the limited amount of training data, we capitalize on the pandemic's asynchronous outbreaks across countries and use a model-agnostic meta-learning based method to transfer knowledge from one country's model to another's. We compare the proposed approach against simple baselines and more traditional forecasting techniques in 4 European countries. Experimental results demonstrate the superiority of our method, highlighting the usefulness of GNNs in epidemiological prediction. Transfer learning provides the best model, highlighting its potential to improve the accuracy of the predictions in case of secondary waves, if data from past/parallel outbreaks is utilized.
\end{abstract}

\section{Introduction}\label{submission}
In late 2019, a highly infectious new virus, SARS-CoV-2, started spreading in Wuhan, China.
In early 2020, the virus had spread to most countries around the world causing the pandemic of the COVID-19 disease.
As of December 7, 2020, a total of 1,532,418 deaths and 66,422,058 cases of COVID-19 were confirmed worldwide\footnote{\url{https://covid19.who.int}}.
During a pandemic, accurately predicting the spread of the infection is of paramount importance to governments and policymakers in order to impose measures to combat the spread of the virus or decide on the allocation of healthcare resources.

Given the severity of the pandemic and the need for accurate forecasting of the disease spread, machine learning, and artificial intelligence approaches have recently started to emerge as a promising methodology to combat COVID-19.
However, it turns out that the nature of the problem is rather challenging \cite{zeroual2020deep}.
In many domains, data admits a natural graph representation.
For instance, to predict the spread of COVID-19, ideally, we would like to have access to the social network of all individuals and to make predictions based on the interactions between them. However, in the absence of such data, we consider a similar problem;  predicting the development of the disease based on mass mobility data, \ie how many people moved from one place to another.
Mobility inside a region can be regarded as a proxy of the interaction i.e. the more people move, the higher the risk of transmission inside the region.
Interestingly, mobility between different regions is known to play a crucial role in the growth of the pandemic, especially for long range travels \cite{colizza2006prediction,soriano2020impact}.
Mobility gives rise to a natural graph representation allowing the application of recent relational learning techniques such as graph neural networks (GNNs).

GNNs have been applied to a wide variety of tasks, including node classification \cite{kipf2017semi}, graph classification \cite{morris2019weisfeiler} 
and text categorization \cite{nikolentzos2020message}.
GNNs capitalize on the concept of message passing, that is, at every iteration, the representation of each vertex is updated based on messages received from its neighbors.
Since GNNs have been successfully applied to several real-world problems, in this paper, we also investigate their effectiveness in forecasting COVID-19. 
We focus on the problem of predicting the number of confirmed COVID-19 cases in each node.
We propose a model that captures both spatial and temporal information, thus combining mobility data with the history of COVID-19 cases.

To investigate if the model can learn the underlying complex dynamics associated with COVID-19, we evaluate it on recent data where we predict the number of new cases.
Our results demonstrate that GNNs on mobility exhibit a substantial potential in predicting the disease spread.
Furthermore, since the availability of data is limited at the start of the outbreak in a country, we employ a transfer learning method based on Model-Agnostic Meta-Learning (MAML) to capitalize on knowledge from other countries' models. Our main contributions are summarized as follows:
\begin{itemize}
    \item We propose a model for learning the spreading of COVID-19 in a country's graph of regions. The model relies on the representational power of GNNs and their capability to encode the underpinnings of the epidemic.
    \item We apply a method based on MAML to transfer a disease spreading model from countries where the outbreak has been stabilized, to another country where the disease is at its early stages. 
    \item We evaluate the proposed approach on data obtained from regions of 4 different countries, namely France, Italy, Spain, and England. We observe that it can indeed surpass the benchmarks and produce useful predictions.
\end{itemize}

The rest of this paper is organized as follows. Section \ref{sec:related_work} provides an overview of the related work and elaborates our contribution. Section \ref{sec:dataset} provides an overview of the dataset and the relationship between mobility and COVID-19 cases.
Section \ref{sec:method} provides a detailed description of the proposed model. Section \ref{sec:experiments} evaluates the proposed model on data from the first COVID-19 wave in 4 EU countries. Finally, section \ref{sec:conclusion} summarizes the work and presents potential future work.

\section{Related Work}\label{sec:related_work}
As mentioned above, many recent studies have leveraged machine learning and artificial intelligence techniques to make predictions about the spread of COVID-19.
For instance, Lorch et al. (\citeyear{lorch2020spatiotemporal}) propose a compartmental SEIR model which is based on a parameterized counting process.
The parameteres of the model are estimated using bayesian optimization.
and was evaluated on regions of Germany and Switzerland.
Flaxman et al. (\citeyear{flaxman2020estimating}) study the effect of major non-pharmaceutical interventions across 11 European countries
with a bayesian model whose parameters are estimated based on the observed deaths in those countries. Their results indicate that the interventions have had a large effect on reducing the spread of the disease.
Time-series based models have also been utilized and will serve as our baselines.
For instance, Chimmula and Zhang (\citeyear{chimmula2020time}) employed an LSTM to predict the number of confirmed COVID-19 cases in Canada, while Kufel (\citeyear{kufel2020arima}) investigated the effectiveness of the ARIMA model in predicting the dynamics of COVID-19 in certain European countries.

A GNN for epidemic forecasting, ColaGNN, was recently developed by Deng et al. (\citeyear{deng2019graph}).
ColaGNN learns a hidden state for each location using an RNN, and then an attention matrix is derived from these representations that captures how locations influence each other.
This matrix forms the graph that is passed on to a GNN to generate the outputs.
This work was evaluated on influenza-like illness (ILI) prediction in US and Japan, without the use of an underlying graph. 
More recent works on predicting COVID-19 using the graph of US counties include a static \cite{kapoor2020examining} and a temporal GNN \cite{gao2020stan}.
The former forms a supergraph using the instances of the mobility graph, where the spatial edges capture county-to-county movement at a specific date, and a county is connected to a number of past instances of itself with temporal edges.
The node features include demographics, number of deaths and recoveries. 
In STAN \cite{gao2020stan} on the other hand, the edges are determined based on demographic similarity and geographical proximity between the counties.
STAN takes advantage of the nature of the pandemic and predicts the parameters of an epidemic simulation model together with the infected and recovered cases, using multiple outputs in the neural network.
These are used to produce long-term predictions based on the simulation and to penalize the original long-term predictions of the model.
The main difference between these approaches and our work lies in the size of the constructed graph and the amount of available data for training.
To be specific, in both these approaches the size of the training data ranges from $50$ to $60$ days.
In our case, this is not feasible as the pandemic is already at its peak before the $30^{\text{th}}$ day, and has already cost too many lives.
This is why we utilize transfer learning to account for the limited training samples in the initial stages of the pandemic.
Moreover, our graphs are relatively small compared to the graph of US counties.
Finally, our open data lacks in many cases the number of recovered cases, deaths and population demographics required for training these models.
This kind of data may not always be available at the regional level for a real-life pandemic, especially for smaller, less developed countries.


Transfer learning for disease prediction has been used in the past in Zou et al. (\citeyear{zou2019transfer}) who have mapped a disease model trained on online google searches obtained from one location, where the virus spreading is available, to another location, where the virus has not spread widely yet. More recently, this approach was utilized in the context of COVID-19 \cite{lampos2020tracking}.
In the context of graph representation learning, transfer learning has only been used to the best of our knowledge for classifying textual documents represented as graphs \cite{lee2017transfer}, for traffic prediction \cite{mallick2020transfer}, for semi-supervised classification \cite{yao2019graph} and for designing GNNs that are robust to adversarial attacks \cite{tang2020transferring}. 

\section{Dataset}\label{sec:dataset}
Facebook has released several datasets in the scope of Data For Good program\footnote{https://dataforgood.fb.com/tools/disease-prevention-maps/} to help researchers better understand the dynamics of the COVID-19 and forecast the spread of the disease  \cite{maas2019facebook}.
We use a dataset that consists of measures of human mobility between administrative NUTS3\footnote{\url{https://en.wikipedia.org/wiki/Category:NUTS_3_statistical_regions_of_the_European_Union}} regions.
The data is collected directly from mobile phones that have the Facebook application installed and the Location History setting enabled.
The raw data contains three recordings per day (\ie midnight, morning and afternoon), indicating the number of people moving from one region to another at that point of day. We compute a single value for each day and each pair of regions by aggregating these three values. We focus on $4$ European countries: Italy, Spain, France and England. 

The number of cases in the different regions of the 4 considered countries were gathered from open data listed in the github page\footnote{\url{https://github.com/geopanag/pandemic_tgnn}} along with the code and the aggregated mobility data. 
An overview of the preprocessed data can be found in Table~\ref{country_table}.
The start date is the earliest date for which we have both mobility data and data related to the number of cases available.
We apply text mining techniques to preprocess and map the regions of the mobility data to those of the open data, as well as quality control for noisy time series where the recordings seemed infeasible.
This led us into neglecting a 2 in Italy, and 10 (including islands) in Spain.
We also do not take into consideration regions that had less then 10 confirmed cases in total throughout the pandemic, which corresponds to 14 regions in Spain and 3 in Italy, as they were luckily not very affected by the pandemic.

\begin{table}[t]
\centering
\begin{sc}
\begin{tabular}{lcccr}
    \toprule
    Country & Time & Regions & Avg new case \\
    \midrule
    Italy  & 24/2-12/5 & 105& 25.65\\
    England & 13/3-12/5 & 129 & 16.7\\
    Spain  & 12/3-12/5 & 35& 61\\
    France & 10/3-12/5 & 81 & 7.5\\ 
    \bottomrule
\end{tabular}
\end{sc}
\caption{Summary of the available data for the 3 considered countries.}
\label{country_table}
\end{table}

\begin{figure}[t]
\centering
\subfigure{
\hspace{-.8cm}
\includegraphics[width=1.3\linewidth]{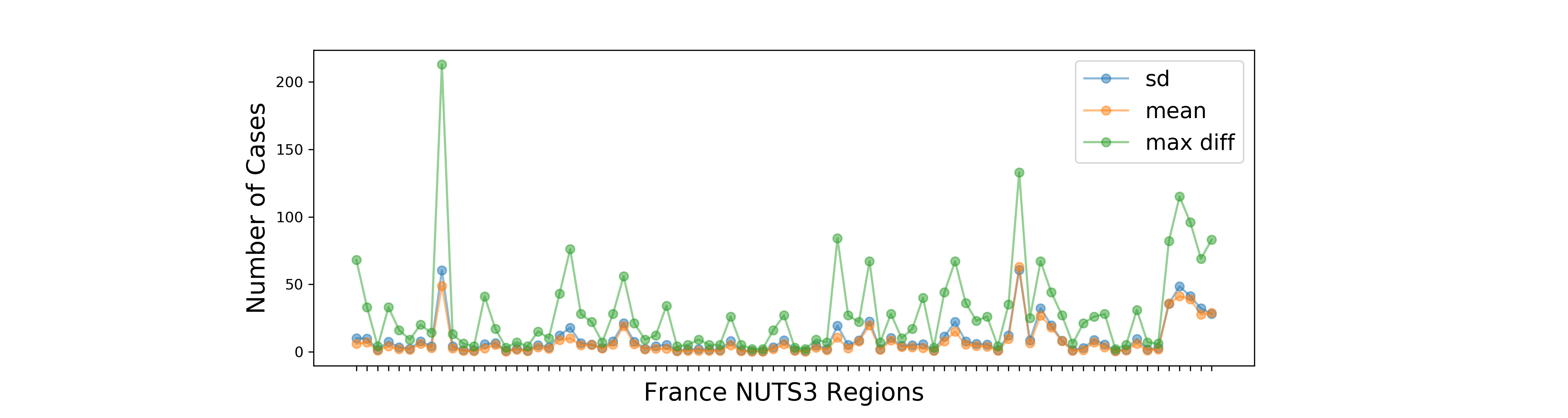}
}
\subfigure{

\hspace{-.8cm}
\includegraphics[width=1.3\linewidth]{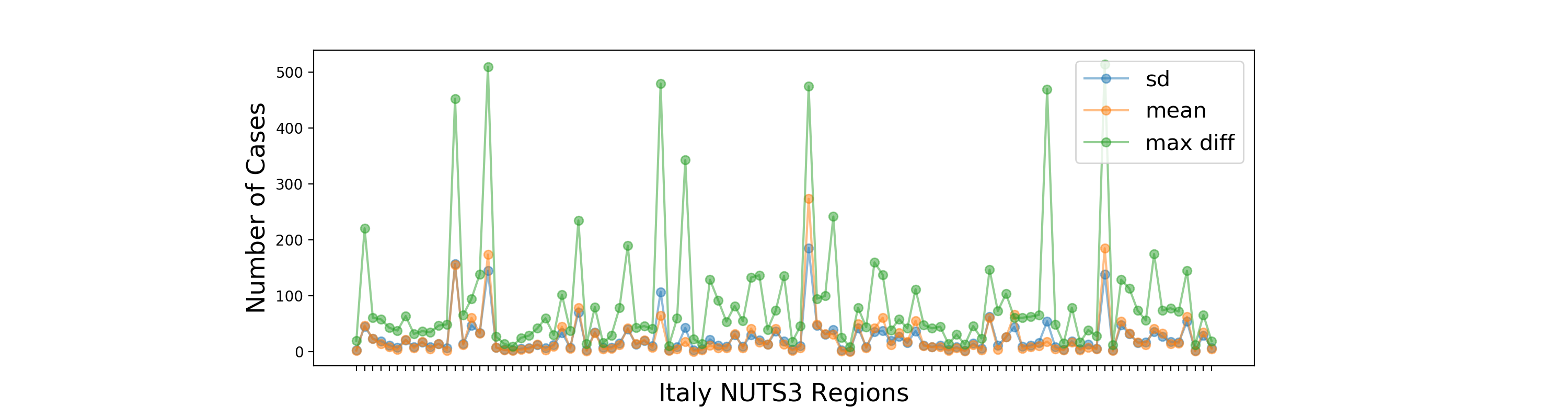}
}

\caption{Mean, standard deviation and maximum difference of confirmed cases per day.}
\label{fig:irregularity}
\end{figure}

We should stress here that in the considered set of data, case reporting is often not consistent, while there are also very large variations in the number of tests performed in each region/country.
This is the main reason behind the large differences in the number of reported cases from day to day, as illustrated in Figure~\ref{fig:irregularity} for the different regions of Italy and France. Specifically, we see that for almost all regions, the maximum difference encountered between consecutive days is multiple times the average value of the time series, signifying the burstiness and the difficulty of predicting exact samples of such time series.

\begin{figure}[t!]

\begin{minipage}{0.05cm}\rotatebox{90}{NUTS3 Regions}\end{minipage}
\begin{minipage}[b]{\columnwidth}
\centering
\subfigure{\includegraphics[width=0.47\linewidth]{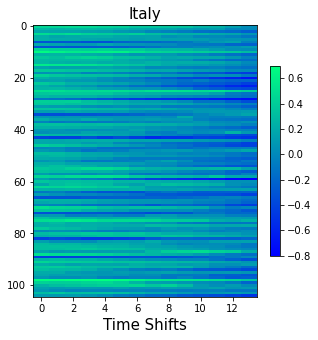}}
\subfigure{\includegraphics[width=0.45\linewidth]{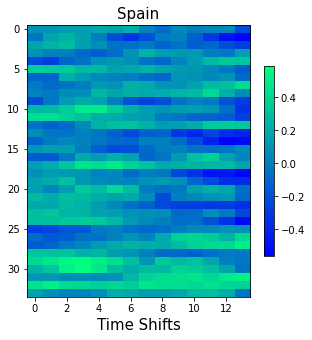}}
\end{minipage}

\begin{minipage}{0.05cm}\rotatebox{90}{}\end{minipage}
\begin{minipage}[b]{\columnwidth}
\centering
\vspace{-1cm}
\subfigure{\includegraphics[width=0.45\linewidth]{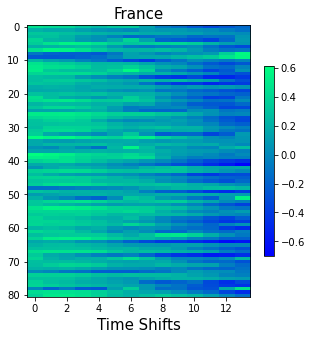}}
\subfigure{\includegraphics[width=0.45\linewidth]{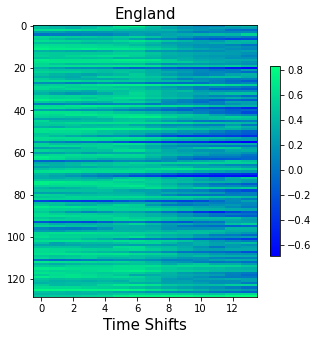}}
\end{minipage}
\caption{Pearson correlation between mobility and number of confirmed cases in the future for examined countries's regions.}
\label{fig:mobility_cases}
\end{figure}

In order to evaluate the relationship between mobility and COVID-19 cases, we compute the pearson correlation for the examined time shifts in forecasting i.e. ranging from 1 to 14 days ahead. Specifically, for a region $u$, its mobility is the total number of people moving in and out of it each day, represented by $m^t$ starting from time 0 to time $t$. The sequence of confirmed COVID-19 cases is represented by $c^t$, and $c^{t+1}$ represents the vector starting from time 1 to $t+1$. So for a shift of 1, $t=T-1$ and $Corr^1 = \frac{ \sum_i (m_i^t-\mu(m_i^t)(c_i^{t+1}-\mu(c_i^{t+1})) }{\sigma(m^t) \sigma(c^{t+1})}$. In Fig. \ref{fig:mobility_cases} we can see the correlation for all regions and all time shifts.
Overall, we can see that for most of the regions, mobility is correlated positively with the short term number of cases and vice versa for the long term ones. Overall this pattern pertains throughout most of the regions with the exception of Spain. Hence we expect mobility to be a usfull predictor.

\section{Methodology}\label{sec:method}
In this section, we present the proposed neural network architecture for predicting the course of the COVID-19 disease.
It should be mentioned that our analysis involves a series of assumptions.
First, we assume that people that use Facebook on their mobile phones with Location History enabled constitute a uniform random sample of the general population.
Second, we assume that the number of cases in a region reported by the authorities is a representative sample of the number of people that have been actually infected by the virus.
Finally, we hypothesize that the more people move from one region to another or within a region, the higher the probability that people in the receiving region are infected by the virus.
This is a well-known observation in the field of epidemics\cite{colizza2006prediction,soriano2020impact}, and motivates the use of a message passing procedure as we delineate below.

\subsection{Graph Construction}
We chose to represent each country as a graph $G=(V,E)$ where $n=|V|$ denotes the number of nodes.
Specifically, given a country, we create a series of graphs, each corresponding to a specific date $t$, \ie $G^{(1)}, \ldots, G^{(T)}$.
A single date's mobility data is transformed into a weighted, directed graph whose vertices represent the NUTS3 regions and edges capture the mobility patterns.
For instance, the weight $w_{v,u}^{(t)}$ of the edge $(v,u)$ from vertex $v$ to vertex $u$ denotes the total number of people that moved from region $v$ to region $u$ at time $t$.
Note that these graphs can also contain self-loops which correspond to the mobility behavior within the regions. 
The mobility between administrative regions $u$ and $v$ at time $t$ forms an edge which, multiplied by the number of cases $c_u^{(t)}$ of region $u$ at time $t$, provides a relative score expressing how many infected individuals might have moved from $u$ to $v$.
To be more specific, let $\mathbf{x}_u^{(t)} = (c_u^{(t-d)}, \ldots, c_u^{(t)})^\top \in \mathbb{R}^d$ be a vector of node attributes, which contains the number of cases for each one of the past $d$ days in region $u$. We use the cases of multiple days instead of just the day before the prediction because case reporting is highly irregular between days, especially in decentralized regions. Intuitively, message passing over this network computes a feature vector for each region with a combined score from all regions, as illustrated below. 
\begin{equation*}
    \mathbf{A}^{(t)} \, \mathbf{X}^{(t)} = 
    \begin{bmatrix}
      w_{1,1}^{(t)} & w_{2,1}^{(t)} & \ldots & w_{n,1}^{(t)} \\[0.3em]
      w_{1,2}^{(t)} & w_{2,2}^{(t)} & \ldots & w_{n,2}^{(t)}  \\[0.3em]
      \vdots & \vdots & \vdots & \vdots \\[0.3em]
      w_{1,n}^{(t)} & w_{2,n}^{(t)} & \ldots & w_{n,n}^{(t)} 
    \end{bmatrix}
    \begin{bmatrix}
      \mathbf{x}_1^{(t)} \\[0.3em]
      \mathbf{x}_2^{(t)} \\[0.3em]
      \vdots \\[0.3em]
      \mathbf{x}_3^{(t)} 
    \end{bmatrix}
    = \\
    \begin{bmatrix}
      \mathbf{z}_1 \\[0.3em]
      \mathbf{z}_2 \\[0.3em]
      \vdots \\[0.3em]
      \mathbf{z}_3 
    \end{bmatrix}
    \label{eq:mp}
\end{equation*}

\begin{figure}
\centering
\begin{tikzpicture}

	\node[circle, draw, thick] (u) {$u$};
	\node[circle, draw, thick, above left=of u] (v) {$v$};
	\node[circle, draw, thick, left=5em of u] (j) {$j$};
	\node[circle, draw, thick, above right=of u] (i) {$i$};
	
	\node[below=0.5cm of u] (eq) {$ Z_u = (x_j a_{j,u} + x_i a_{i,u}+ x_v a_{v,u}) + x_u a_{u,u} $};
	
    \draw[-stealth, blue, thick, decoration={snake, pre length=0.01mm, segment length=1mm, amplitude=0.3mm, post length=1mm}, decorate] (v.-45) -- node[sloped, above, black] {$a_{vu}$} (u.135);
    \draw[-stealth, blue, thick, decoration={snake, pre length=0.01mm, segment length=1mm, amplitude=0.3mm, post length=1mm}, decorate] (i.-140) -- node[sloped, above, black] {$a_{iu}$} (u.60);
    \draw[-stealth, blue, thick, decoration={snake, pre length=0.01mm, segment length=1mm, amplitude=0.3mm, post length=1mm}, decorate] (j.0) -- node[sloped, above, black] {$a_{ju}$} (u.170);
    \draw[-stealth, blue, thick, decoration={snake, pre length=0.01mm, segment length=1mm, amplitude=0.3mm, post length=1mm}, decorate] (u) to[looseness=5, out=.-45, in=  +30] node[sloped, below, black] {$a_{uu}$} (u);
    
    \draw[ -stealth, black, opacity=0.5, ultra thick] (u.-90) -- (eq);
\end{tikzpicture}
\caption{Example of the message passing.}
\label{fig:message}
\end{figure}
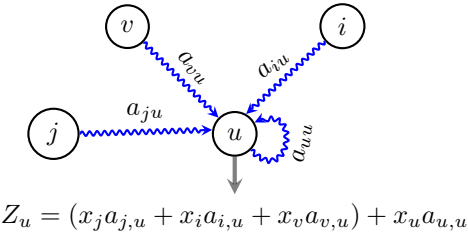
where $\mathbf{A}^{(t)}$ is the adjacency matrix of $G^{(t)}$ and $\mathbf{X}^{(t)}$ is a matrix whose rows contain the attributes of the different regions.
In this case, $\mathbf{z}_u \in \mathbb{R}^d$ is a vector that combines the mobility within and towards region $u$ with the number of reported cases both in $u$ and in all the other regions. Here, we would like to stress the importance of the mobility patterns $w_{u,u}$ within a region $u$ which correspond to good indicators of the evolution of the disease, especially during lockdown periods. To visualize concretely how the representations are extracted from the message passing, Fig. (\ref{fig:message}) contains a toy example with a region $u$ receiving $a$ people from different regions, $x$ containing a vector of past cases in that region.
$Z_u \in \mathbb{R}^d$ represents an estimate of the number of new latent cases in $u$, and is broken down to the cases received from other regions (inside the parenthesis) and the new cases caused due to mobility inside $u$ ($x_u a_{uu}$).

\subsection{Models}
To model the dynamics of the spreading process, we use two instances of a well-known family of GNNs, known as message passing neural networks (MPNNs) \cite{gilmer2017neural}.
These networks consist of a series of neighborhood aggregation layers.
Each layer uses the graph structure and the node feature vectors from the previous layer to generate new representations for the nodes.

\paragraph{MPNN.}
To update the representations of the vertices of each of the input graphs, we use the following neighborhood aggregation scheme:
\begin{equation*}
    \mathbf{H}^{i+1} = f(\tilde{\mathbf{A}} \, \mathbf{H}^i \, \mathbf{W}^{i+1})
\end{equation*}
where $\mathbf{H}^i$ is a matrix that contains the node representations of the previous layer, with $\mathbf{H}^0=\mathbf{X}$, $\mathbf{W}^i$ is the matrix of trainable parameters of layer $i$, and $f$ is a non-linear activation function such as ReLU. 
Following Kipf and Welling (\citeyear{kipf2017semi}), we normalize the adjacency matrix $\mathbf{A}$ such that the sum of the weights of the incoming edges of each node is equal to $1$.
Note that for simplicity of notation, we have omitted the time index.
The above model is in fact applied to all the input graphs $G^{(1)}, \ldots, G^{(T)}$ separately.
Given a model with $K$ neighborhood aggregation layers, the matrices $\tilde{\mathbf{A}}$ and $\mathbf{H}^0, \ldots, \mathbf{H}^K$ are specific to a single graph, while the weight matrices $\mathbf{W}^1, \ldots, \mathbf{W}^K$ are shared across all graphs.
\begin{figure}[t]
\centering
\includegraphics[width=1\columnwidth]{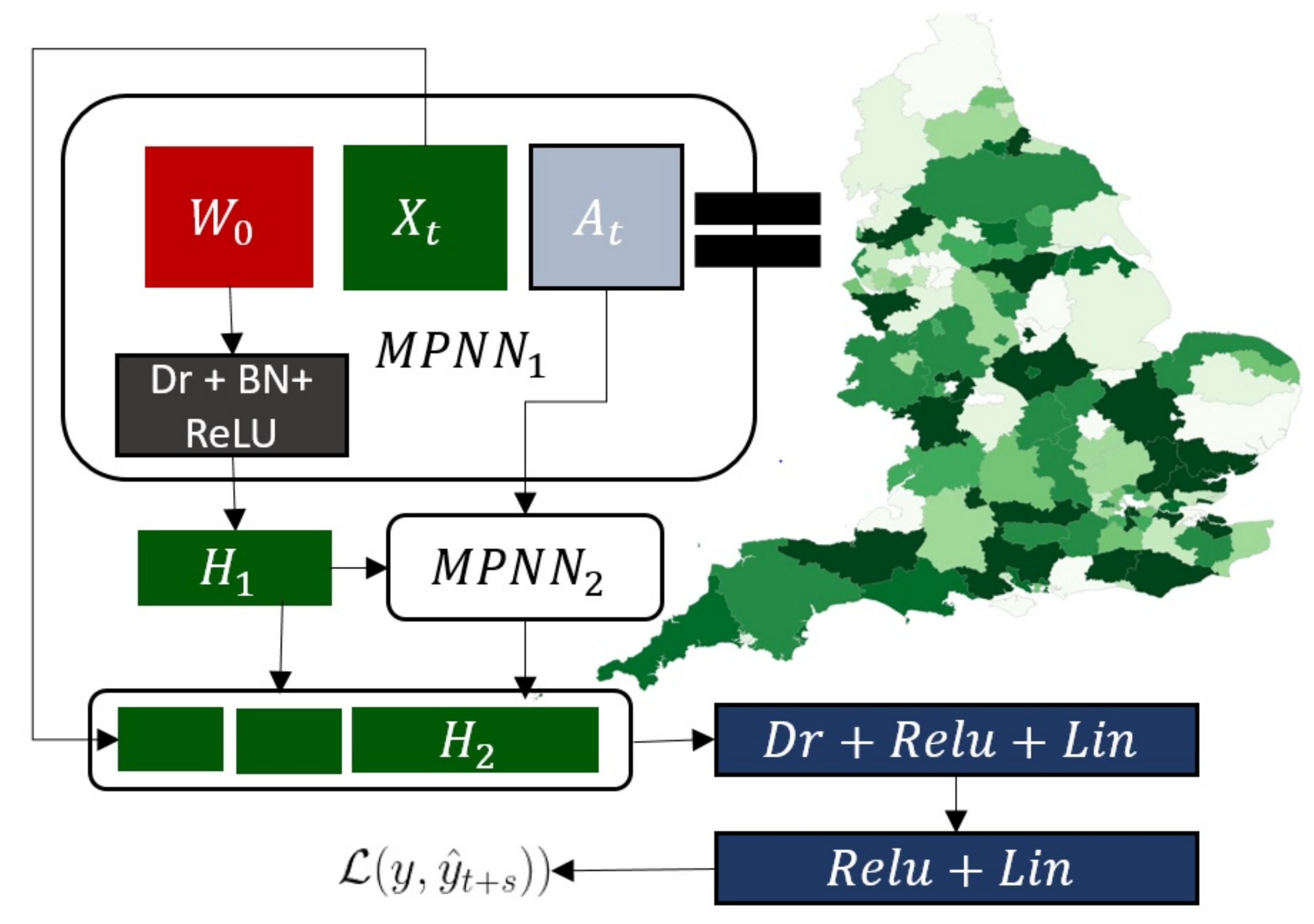}
\caption{Overview of the proposed \textbf{MPNN} architecture.}
\label{fig:method_scheme}
\end{figure}
Typically, an MPNN contains $K$ neighborhood aggregation layers.
As the number of neighborhood aggregation layers increases, the final node features capture more and more global information.
However, retaining local, intermediary information might be useful as well.
Thus, we concatenate the matrices $\mathbf{H}^0,\mathbf{H}^1,\mathbf{H}^2,\ldots,\mathbf{H}^K$ horizontally, \ie $\mathbf{H} = \textsc{\small CONCAT}(\mathbf{H}^0,\mathbf{H}^1,\mathbf{H}^2,\ldots,\mathbf{H}^K)$, and the rows of the emerging matrix $\mathbf{H}$ can be regarded as vertex representations that encode multi-scale structural information, including the initial features of the node.
In other words, we utilize skip connections from each layer to the output layer which consists of a sequence of fully-connected layers.
Note that we apply the ReLU function to the output of the network since the number of new cases is a nonnegative integer.
We choose the mean squared error as our loss function as shown below:
\begin{equation}
    \mathcal{L} = \frac{1}{nT}\sum_{t=1}^T\sum_{v \in V} \big(y_v^{(t+1)}-\hat{y}_v^{(t+1)} \big)^2 
    \label{eq:loss}
\end{equation}
where $y_v^{(t+1)}$ denotes the reported number of cases for region $v$ at day $t+1$ and $\hat{y}_v^{(t+1)}$ denotes the predicted number of cases.
An overview of the \textbf{MPNN} is given in Figure~\ref{fig:method_scheme}.

\paragraph{MPNN+LSTM.}
In order to take advantage of the temporal correlation between the target and the confirmed cases in the past, we build a time-series version of our model using the different snapshots of the mobility graph.
Given a sequence of graphs $G^{(1)}, G^{(2)}, \ldots, G^{(T)}$ that correspond to a sequence of dates, we utilize an MPNN at each time step, to obtain a sequence of representations $\mathbf{H}^{(1)},\mathbf{H}^{(2)},\ldots,\mathbf{H}^{(T)}$.
These representations are then fed into a Long-Short Term Memory network (LSTM) \cite{graves2014towards} which can capture the long-range temporal dependencies in time series.
We expect the hidden states of an LSTM to capture the spreading dynamics based on the mobility information encoded into the node representations.
We use a stack of two LSTM layers.
The new representations of the regions correspond to the hidden state of the last time step of the second LSTM layer.
These representations are then passed on to an output layer similar to the \textbf{MPNN}, along with the initial features for each time step.
Note that this model resembles other attempts to spatio-temporal prediction, but instead of convolutional \cite{yao2018deep}, it employs message passing layers, similar to \cite{seo2018structured}. 
\begin{figure}[h]
    \includegraphics[width=.5\textwidth]{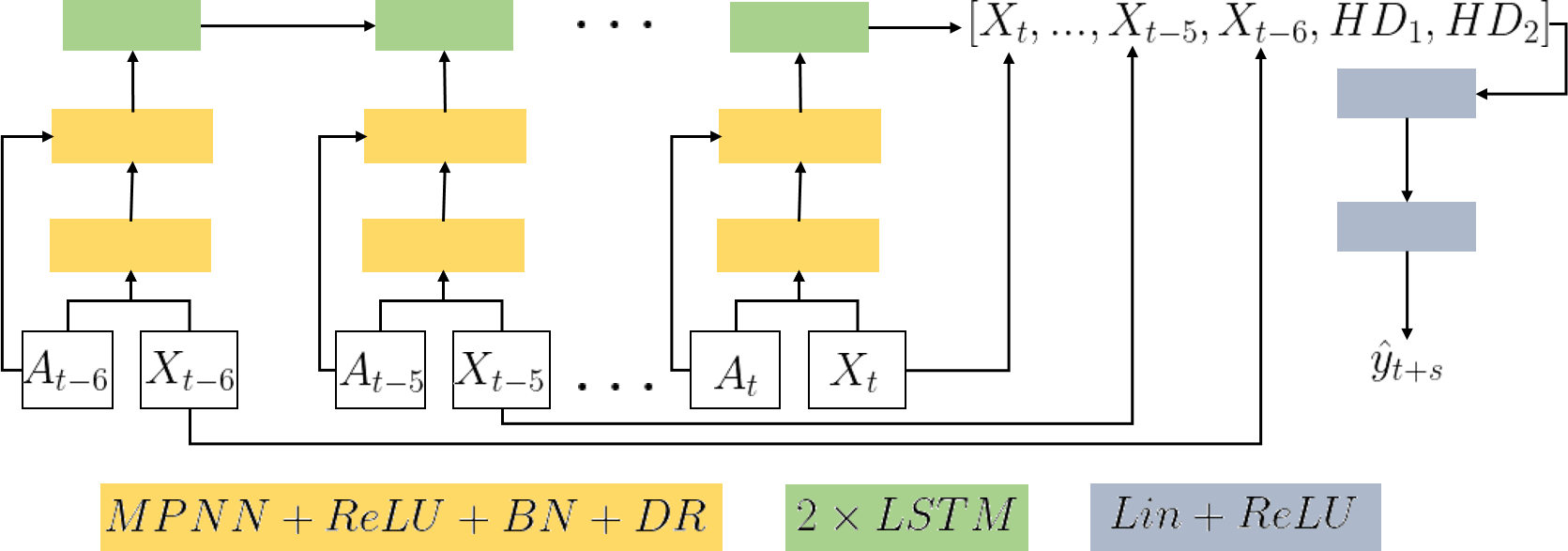}
    \caption{Overview of the MPNN LSTM architecture.}
\end{figure}

\paragraph{MPNN+TL.}
Note that the different countries were hit by the pandemic at different times.
Indeed, there are cases where once the epidemic starts developing in one country, it has already stabilized in another.
Furthermore, a new wave of COVID-19 is very likely to share fundamental characteristics with the previous ones, as it is the same virus.
This additional information may prove rather important in case of insufficient training data.
In our setting, as discussed in subsection~\ref{sec:experimental_setup}, the model starts predicting as early as the $15^{\text{th}}$ day of the dataset.
In such a scenario, the model has access to a few samples to learn from (split into validation and training sets), while it is used to make predictions for as far as $14$ days ahead.
Given the inherent need for data in neural networks, this setting is rather challenging.
Moreover, our intuition is that a model trained in the whole cycle or an advanced stage of the epidemic can capture patterns of its different phases, which is missing from a new model working in a country at the start of its infection.

To incorporate past knowledge from models running in other countries, we separate our data into tasks and propose an adaptation of MAML \cite{finn2017model}.
Lets initially assume that the Meta Train set $M_{\text{tr}}= \{ D^{(1)},\ldots, D^{(p)} \}$, corresponds to the data sets of $p$ countries  that we can use to obtain a set of parameters $\theta$.
The learnt parameters can then be employed to initialize the model for the country left out in the Meta Test $M_{\text{te}}$.
In reality, each dataset $D^{(k)}, k \in \{1,\ldots,p\}$ is divided into subtasks itself.
More specifically, each country has different training sets of increasing size (as the train days increase) as well as shorter- and longer-term targets (next day, two days ahead and so on). For each combination of these two, we train a different model. 
Hence, the set of tasks for a country $k$ is $D^{(k)} = \Big\{ \big(Tr_{i,j}^{(k)},Te_{i,j}^{(k)} \big) : 14 \geq i \geq T_{\text{max}}, 1 \geq j \geq dt \Big\}$ where $\big(Tr_{i,j}^{(k)},Te_{i,j}^{(k)}\big)$ is a dataset (train and test set) associated with country $k$ where the train set comprises of the first $i$ days of the data and the task is to predict the number of cases in the $j$-th day ahead.

The set of parameters $\theta$ corresponds to the weight matrices and biases of all layers in the \textbf{MPNN} model. As mentioned above, in MAML, $\theta$ is randomly initialized and underoges gradient descent steps during the metatrain phase.
The algorithm is shown in Algorithm~\ref{alg:mpnn_maml}.
In each task, we minimize the loss on the task's train set towards a task-specific $\theta_t$, as shown in Equation~(\ref{eq:theta_j}) and on lines 3-5 of the algorithm.
Then, we use the emerging $\theta_t$ to compute the gradient with respect to $\theta$ in the task's test set as illustrated in Equation~(\ref{eq:theta}) below and on line 6.
This gradient is normalized by the total number of tasks in the set to refrain from taking too big steps.

\begin{align}
  \theta_t &= \theta - \alpha \nabla_{\theta}\mathcal{L} \big( f_{\theta}( Tr^{(k)}_{i,j}) \big) 
  \label{eq:theta_j} \\
  \theta &= \theta- \alpha_{m} \nabla_{\theta}\mathcal{L} \big( f_{\theta_t}( Te^{(k)}_{i,j} ) \big) 
  \label{eq:theta}
\end{align}
The standard update includes the gradient of $\theta_t$ and that of $\theta$, which is in fact the hessian matrix, as shown in Equation~(\ref{eq:gradient}).
\begin{equation}
    \frac{\partial \mathcal{L}_{T}(f_{\theta_t}({Te^{(k)}_{i,j}}))}{\partial \theta } = \nabla_{\theta_t}\mathcal{L}(f_{\theta_t}(Te^{(k)}_{i,j}))(I - a\nabla^2_{\theta}\mathcal{L}f_{\theta}(Tr^{(k)}_{i,j})) 
    \label{eq:gradient}
\end{equation}
We are dropping the term that contains the hessian, as it was shown to have insignificant contribution in practice \cite{finn2017model}, possibly due to the vanishing gradient. 
Finally, we train $\theta$ on the train set of $M_{\text{te}}$ and test on its test set (lines 7-10 and 11 respectively).

\begin{algorithm}[!h]
\caption{ \textsc{MPNN+TL}}
\label{alg:mpnn_maml}
\textbf{Input:} $M_{\text{tr}}$, $M_{\text{te}}$, $\alpha$ , $\alpha_m$, $n\_epochs$\\
\textbf{Output:} $\theta$
\begin{algorithmic}[1]
\State Initialize $\theta$ randomly
\For{$D \in M_{\text{tr}}$} 
\For{$(Tr,Te) \in D$} 
\For{Batch $b \in Tr$} 
\State  $\theta_t = \theta - \alpha \nabla_{\theta}\mathcal{L}( {f_{\theta}(b})) $
\EndFor
\State  $\theta = \theta - \alpha_{m} \nabla_{\theta}\mathcal{L} ( {f_{\theta_t}(Te) }) /|M_{\text{tr}}|$
\EndFor
\EndFor
\For{$(Tr,Te) \in M_{\text{te}}$}
\For{Epoch $e \in n\_epochs$}
\For{Batch $b \in Tr$} 
\State  $\theta = \theta - \alpha \nabla_{\theta}\mathcal{L} ({f_{\theta}(b)}) $
\EndFor
\EndFor
\State $error += E( {f_{\theta}(Te) })$
\EndFor
\State \textbf{return} $error/|M_{\text{te}}|$
\end{algorithmic}
\end{algorithm}
Note that in the Algorithm~\ref{alg:mpnn_maml} , $E$ is our error function (\ie Equation~(\ref{eq:error})), $\mathcal{L}$ is the loss function defined in Equation~(\ref{eq:loss}), and $f$ is an \textbf{MPNN}.


\section{Experiments}\label{sec:experiments}
In this section, we first describe the experimental setting and the baselines used for comparison.
We last report on the performance of the proposed models and the baselines.

\subsection{Experimental setup}\label{sec:experimental_setup}

In our experiments, we train the models using data from day $1$ to day $T$, and then use the model to predict the number of cases for each one of the next $dt$ days (\ie from day $T+1$ to day $T+dt$).
We are interested in evaluating the effectiveness of the model in short-, mid- and long-term predictions.
Therefore, we set $dt$ equal to $14$.
We expect the short-term predictions (\ie small values of $dt$) to be more accurate than the long-term predictions (\ie large values of $dt$).
Note that we train a different model to predict the number of cases for days $T+i$ and $T+j$ where $i,j>0$ and $i \neq j$.
Therefore, each model focuses on predicting the number of cases after a fixed time horizon, ranging from $1$ day to $14$ days.
With regards to the value of $T$, it is initially set equal to $14$ and is gradually increased (one day at a time).
Therefore, the size of the training set increases as time progresses.
Note that a different model is trained for each value of $T$.
Furthermore, for each value of $T$, to identify the best model, we build a validation set which contains the samples corresponding to days $T-1$, $T-3$, $T-5$, $T-7$ and $T-9$, such that the training and validation sets have no overlap with the test set.

With regards to the hyperparameters of the \textbf{MPNN}, we train the models for a maximum of $500$ epochs with early stopping after $50$ epochs of patience.
Early stopping starts to occur from the $100^{\text{th}}$ epoch and onward.
We set the batch size to $8$.
We use the Adam optimizer with a learning rate of $10^{-3}$.
We set the number of hidden units of the neighborhood aggregation layers to $64$.
Batch normalization and dropout are applied to the output of every neighborhood aggregation layer, with a dropout ratio of $0.5$.
We store the model that achieved the highest validation accuracy in the disk and then retrieve it to make predictions about the test samples.
For the \textbf{MPNN+LSTM} model, the dimensionality of the hidden states of the LSTMs is set equal to $64$.
All the models are implemented with Pytorch \cite{paszke2019pytorch}.
We evaluate the performance of a model by comparing the predicted total number of cases in each region versus the corresponding ground truth, throughout the test set:
\begin{equation}
    \text{error} = \frac{1}{n\,dt} \, \sum_{t=T+1}^{T+dt} \, \sum_{v \in V} |\hat{y}_v^{(t)}-y_v^{(t)}|
    \label{eq:error}
\end{equation}

\begin{figure}%
\centering
\subfigure{
\label{figure}%
\includegraphics[width=1\columnwidth]{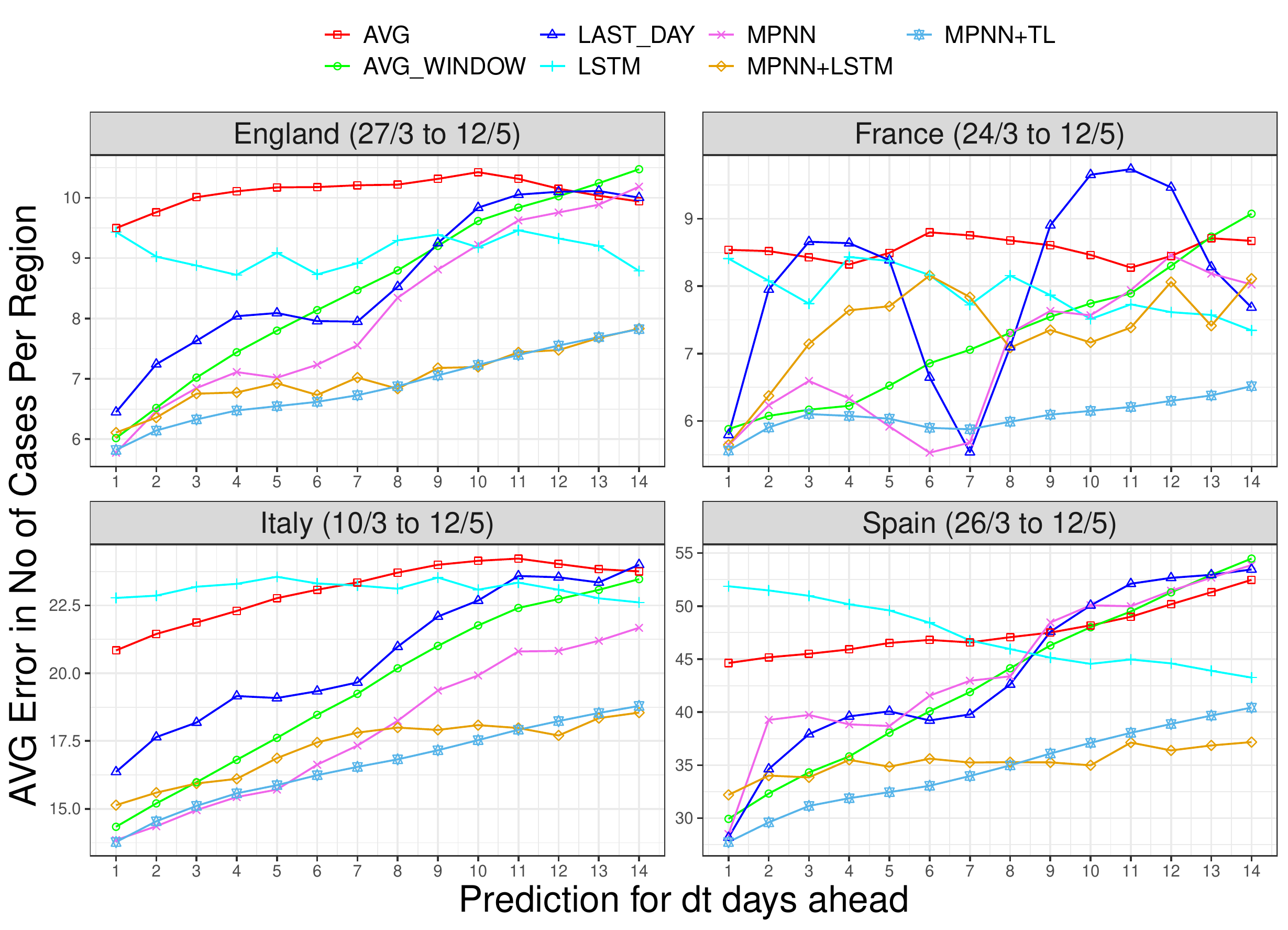}}
\caption{Average number of cases lost per region for each target shift. PROPHET and ARIMA are omitted and shown in the table, because they  effected the legibility of the plot.}
\label{fig:target_error}%
\end{figure}

\begin{table*}[t]
    \centering
    \def\arraystretch{1.1}
    \resizebox{\textwidth}{!}{ 
        \begin{tabular}{|l|rrrr|rrrr|rrrr|} \hline
        \multirow{2}{*}{Model} & \multicolumn{4}{c|}{Up to next 3 Days} & \multicolumn{4}{c|}{Up to next 7 Days} &\multicolumn{4}{c|}{Up to next 14 Days}  \\ \cline{2-13}
        & England &France & Italy & Spain & England &France & Italy & Spain & England &France & Italy & Spain \\
        \hline
        AVG & 9.75 & 8.50 & 21.38 & 45.10 & 9.99 & 8.55 & 22.23 & 45.87& 10.09 & 8.55 & 23.09 & 47.63  \\ 
        LAST\_DAY & 7.11 & 7.47 & 17.40 & 33.58 & 7.62 & 7.37 & 18.49 & 37.06 & 8.66 & 8.03 & 20.69 & 43.63 \\ 
        AVG\_WINDOW & 6.52 & 6.04 & 15.17 & 32.19 & 7.34 & 6.40 & 16.81 & 36.06 & 8.54 & 7.24 & 19.45 & 42.79\\ 
        LSTM & 9.11 & 8.08 & 22.94 & 51.44 & 8.97 & 8.13 & 23.17 & 49.89&  9.10 & 7.91 & 23.12 & 47.26   \\ 
        ARIMA &  13.77 & 10.72 & 35.28 & 40.49 & 14.55 & 10.53 & 37.23 & 41.64 & 15.65 & 10.91 & 39.65 & 46.22 \\ 
        PROPHET &10.58 & 10.34 & 24.86 & 54.76 & 12.25 & 11.56 & 27.39 & 62.16& 16.24 & 14.61 & 33.07 & 79.42 \\
        \hline
        \textbf{TL\_BASE} & 9.65 & 7.67 & 19.12 & 42.25 & 12.30 & 9.21 & 23.44 & 52.29 & 13.48 & 12.27 & 24.89 & 59.68\\
        \textbf{MPNN}  & 6.36 & 6.16 & 14.39 & 35.83 & 6.86 & 5.99 & 15.47 & 38.51 & 8.13 & 6.93 & 17.88 & 44.25\\ 
        \textbf{MPNN+LSTM} & 6.41 & 6.39 & 15.56 & 33.35 & 6.67 & 7.21 & 16.41 & 34.47 & 7.02 & 7.36 & 17.25 & 35.31 \\ 
        \textbf{MPNN+TL} & \textbf{6.05} & \textbf{5.83} & \textbf{14.08} & \textbf{29.61} & \textbf{6.33} & \textbf{5.90} & \textbf{14.61} & \textbf{31.55} &\textbf{6.84} & \textbf{6.13} & \textbf{16.69} & \textbf{34.65}\\
        \hline
        \end{tabular}
    }
    \caption{Average error for $dt=1-3,1-7$ and $1-14$, in number of cases per region.}
    \label{tab:results}
\end{table*}

\subsection{Baselines}
We compare the proposed models against the following baselines and benchmark methods, which have been applied to the problem of COVID-19 forecasting:
\begin{itemize}
\item AVG: The average number of cases for the specific region up to the time of the test day.
\item AVG\_WINDOW: The average number of cases in the past $d$ for the specific region where $d$ is the size of the window.
\item LAST\_DAY: The number of cases in the previous days is the prediction for the next days.
\item LSTM \cite{chimmula2020time}: A two-layer LSTM that takes as input the sequence of new cases in a region for the previous week. 
\item ARIMA \cite{kufel2020arima}: A simple autoregressive moving average model where the input is the whole time-series of the region up to before the testing day.
\item PROPHET \cite{mahmud2020bangladesh}: A forecasting model for various types of time series\footnote{\url{https://github.com/facebook/prophet}}.The input is similar to ARIMA.
\item \textbf{TL\_BASE}: An MPNN that is trained on all data from the three countries and the train set of the fourth (concatenated), and tested on the test set of the fourth. This serves as a baseline to quantify the usefulness of \textbf{MPNN+TL}.
\end{itemize}

We should note here that since we rely solely on the number of confirmed cases, we can not utilize models that work with recovery, deaths and policies, such as SEIR. That said, a simple approach is to run SI at every given $T$ with a parameter $\beta$ taken from the COVID-19 literature, along with the number of infected people at $T$ and the population. In some preliminary experiments, however,  this provided errors in a different scale then the ones mentioned here, similar to Gao et al. (\citeyear{gao2020stan}), which is why we have not experimented further.

\subsection{Results and Discussion}
The average error per region for each one of the next $14$ days is illustrated in Figure~\ref{fig:target_error}.
We observe that in all cases, the proposed models yield lower average errors compared to those of the baselines.
Among the three variants, \textbf{MPNN+TL} is the best-performing model.
It initially outperforms the other approaches by a small margin that increases further after the second day. 
Even simple baselines can be competitive at predicting the next day's number of cases since proximal samples for the same region from the same phases of the pandemic tend to have a similar number of cases.
However, a prediction that goes deeper in time requires the identification of more persistent patterns.
In the case of our model, as mentioned above, we aim to capture unregistered cases moving from one region to the other or spreading the disease in their new region. 
These cases would inevitably take a few days to appear, due to the delay of symptoms associated with COVID-19.
This is why \textbf{MPNN} performs well throughout the 14-days window.
The results also demonstrate the benefit of transfer learning techniques since \textbf{MPNN+TL} outperforms \textbf{MPNN} and its baseline  \textbf{TL\_BASE} in all cases.
We expect \textbf{MPNN} to perform similar towards the end of the dataset, when the training of both models has become similar due to the number of epochs and.
The main difference occurs when $T$ is small, where the training samples are scarce and \textbf{MPNN} is unable to capture the underlying dynamics. One way to see this is again the accuracy of \textbf{MPNN+TL} in the long term predictions. Due to the size of the prediction window, long term tasks have diminished train set, meaning if the task is to predict t+14 and the set ends at day 60, t+14  training will stop at day 46 while the t+1 will stop at 59.  Thus \textbf{MPNN} performs similarly at the short-term predictions but fails compared to \textbf{MPNN+TL} in the long term.

Note that for clarity of illustration, we chose not to visualize the performance of PROPHET and ARIMA in Figure~\ref{fig:target_error} as their error was distorting the plot.
However, we present in Table~\ref{tab:results} the average error for the predictions where $dt$ takes three values: $3$, $7$ and $14$.
Overall, it is clear that the time series methods (\ie LSTM, PROPHET, and ARIMA) and the temporal variant of our method \textbf{MPNN+LSTM} yield quite inaccurate predictions. 
Apart from the inherent difficulty of learning with time-series data, which we analyze further below, this might also happen because of the nature of the epidemic curve.
Specifically, sequential models, that are trained with values that tend to increase, are impossible to predict decreasing or stable values.
For instance, in our dataset, once the models have enough samples to learn from, there is a transition in the phase of the epidemic due to lockdown measures.
The same applies when the epidemic starts to recede at the start of May.

\begin{figure}[t]
\centering
\subfigure{
\includegraphics[height=1.1in]{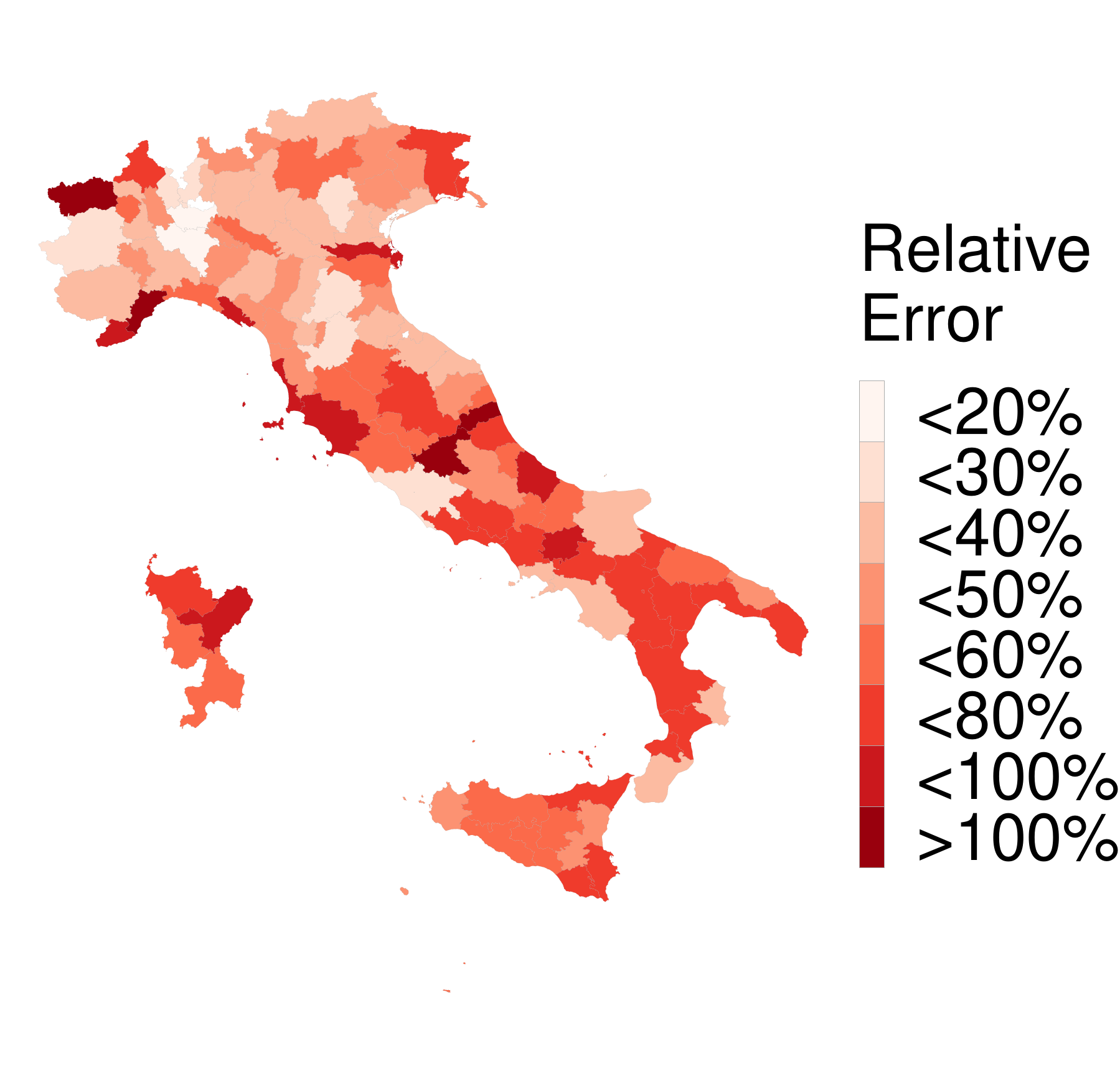}}
\subfigure{
\includegraphics[height=1.1in]{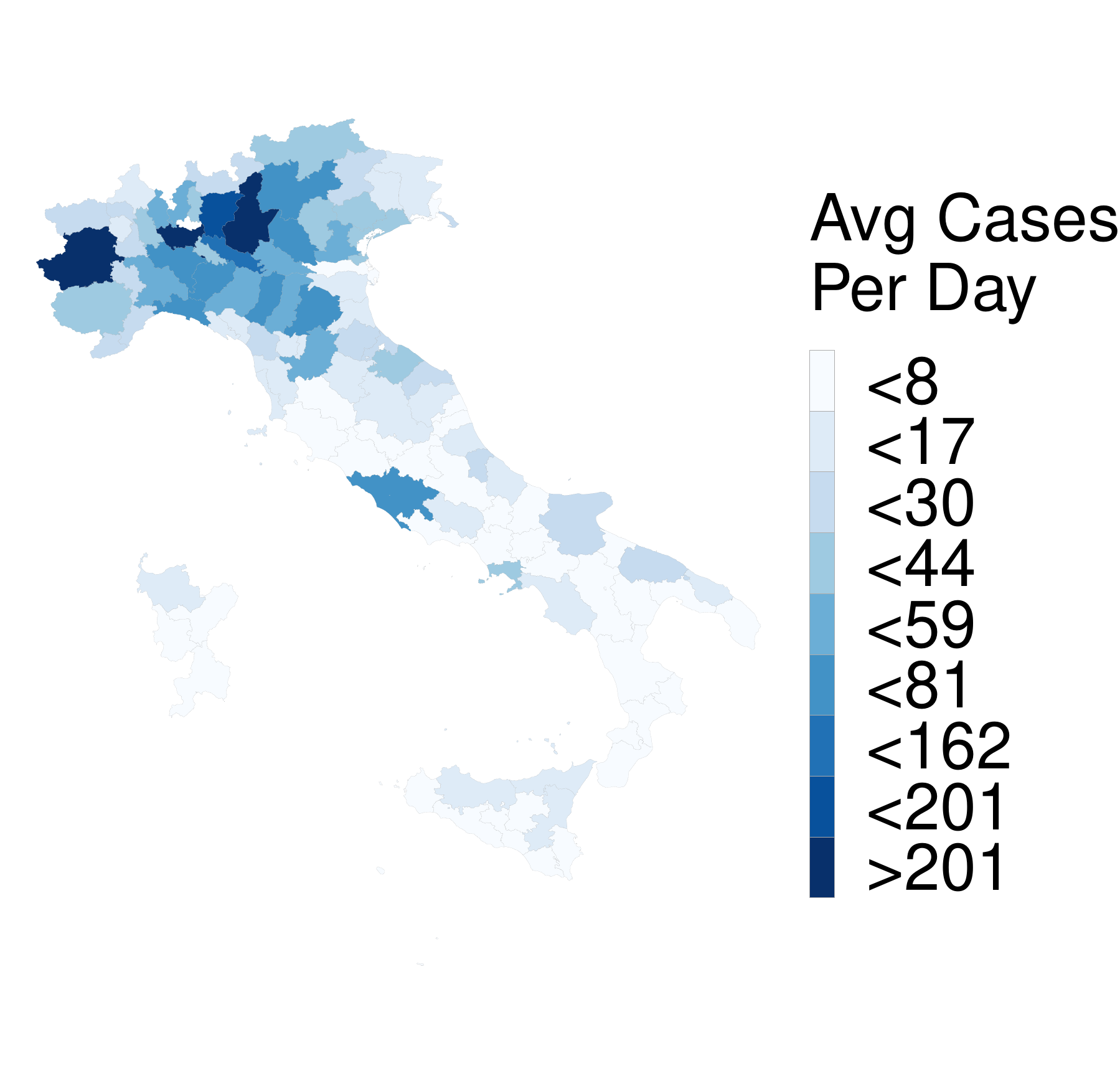}
}
\subfigure{
\includegraphics[height=1.1in]{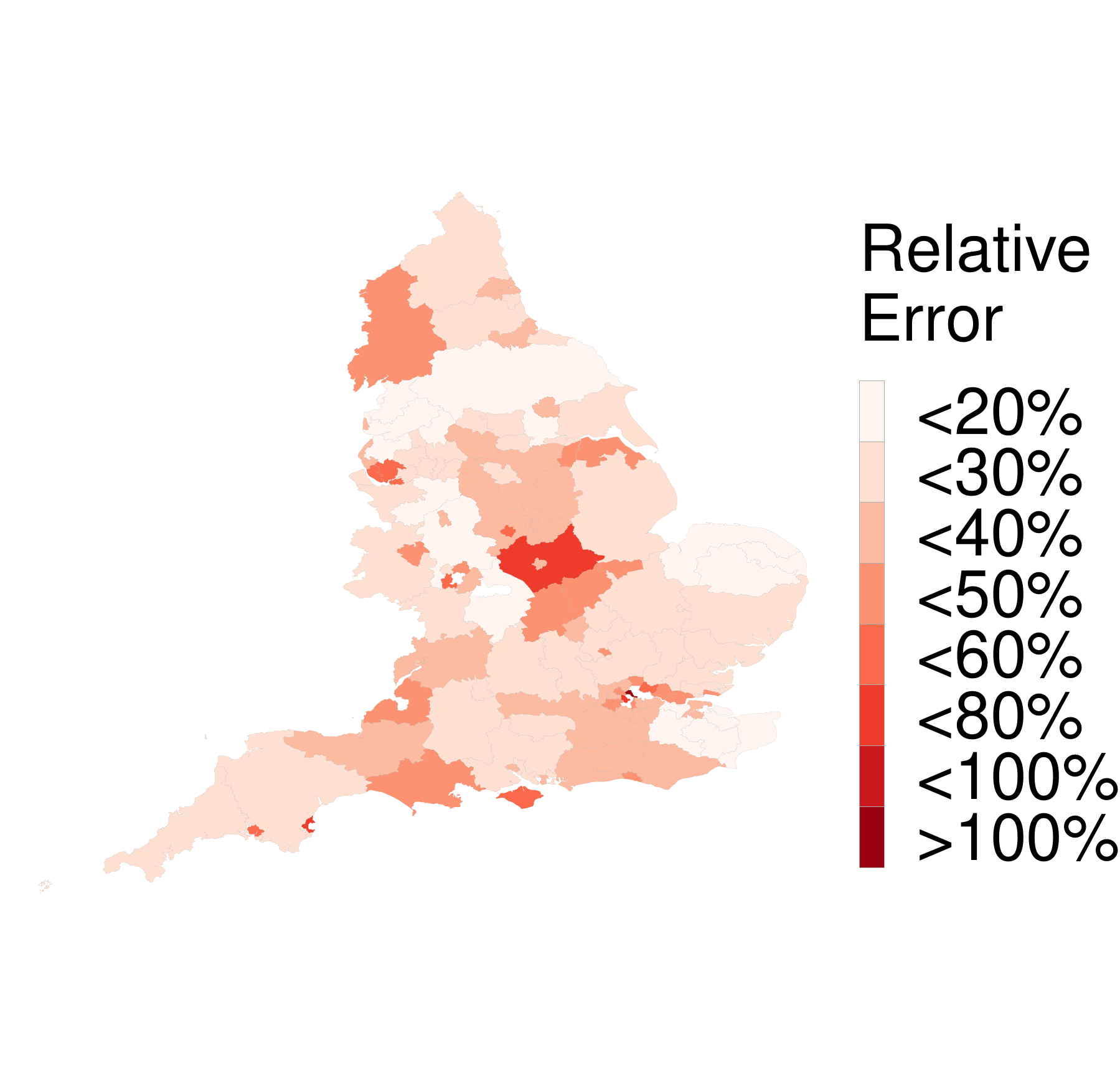}}
\subfigure{
\includegraphics[height=1.1in]{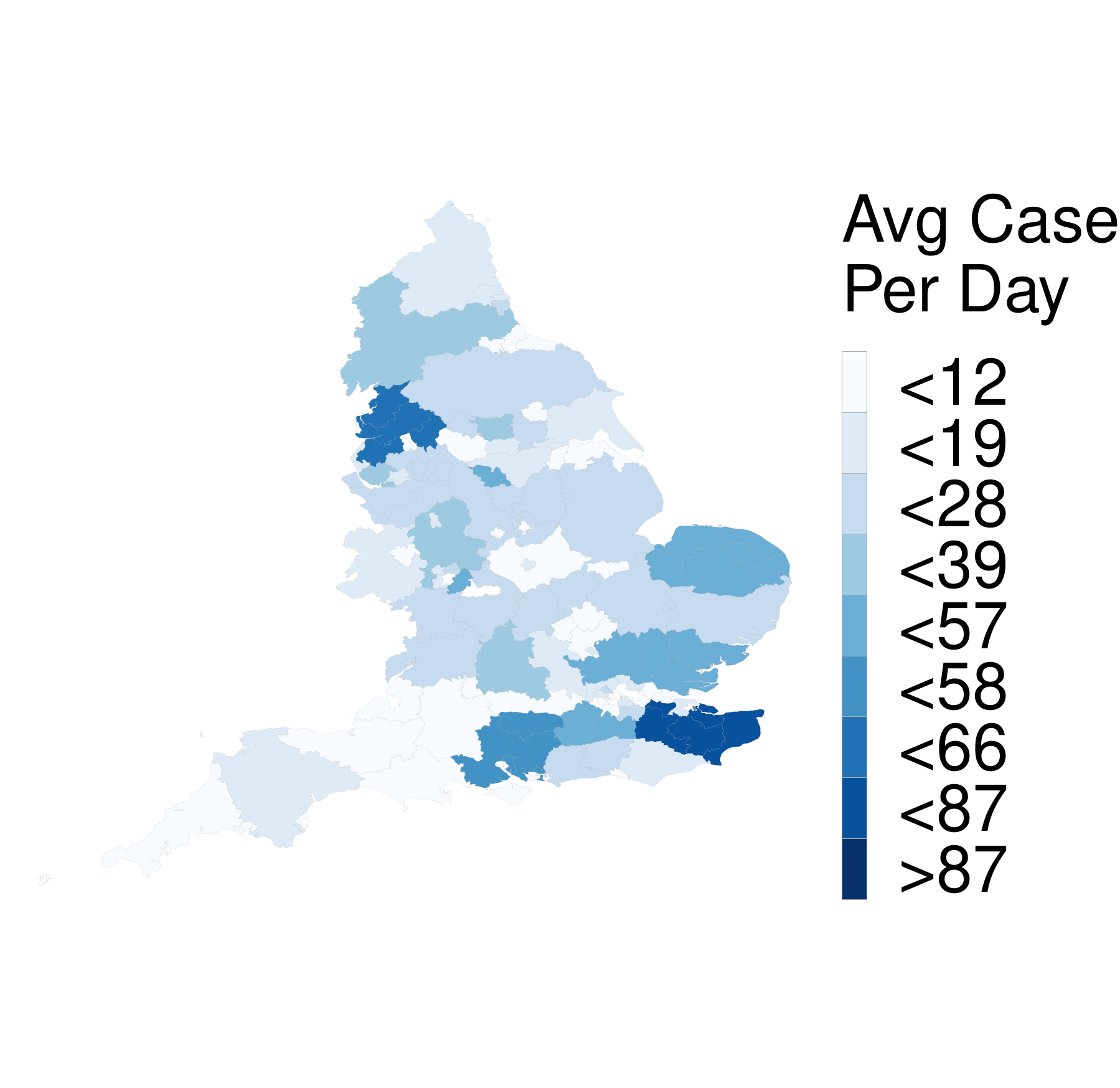}
}
\subfigure{
\includegraphics[height=1.1in]{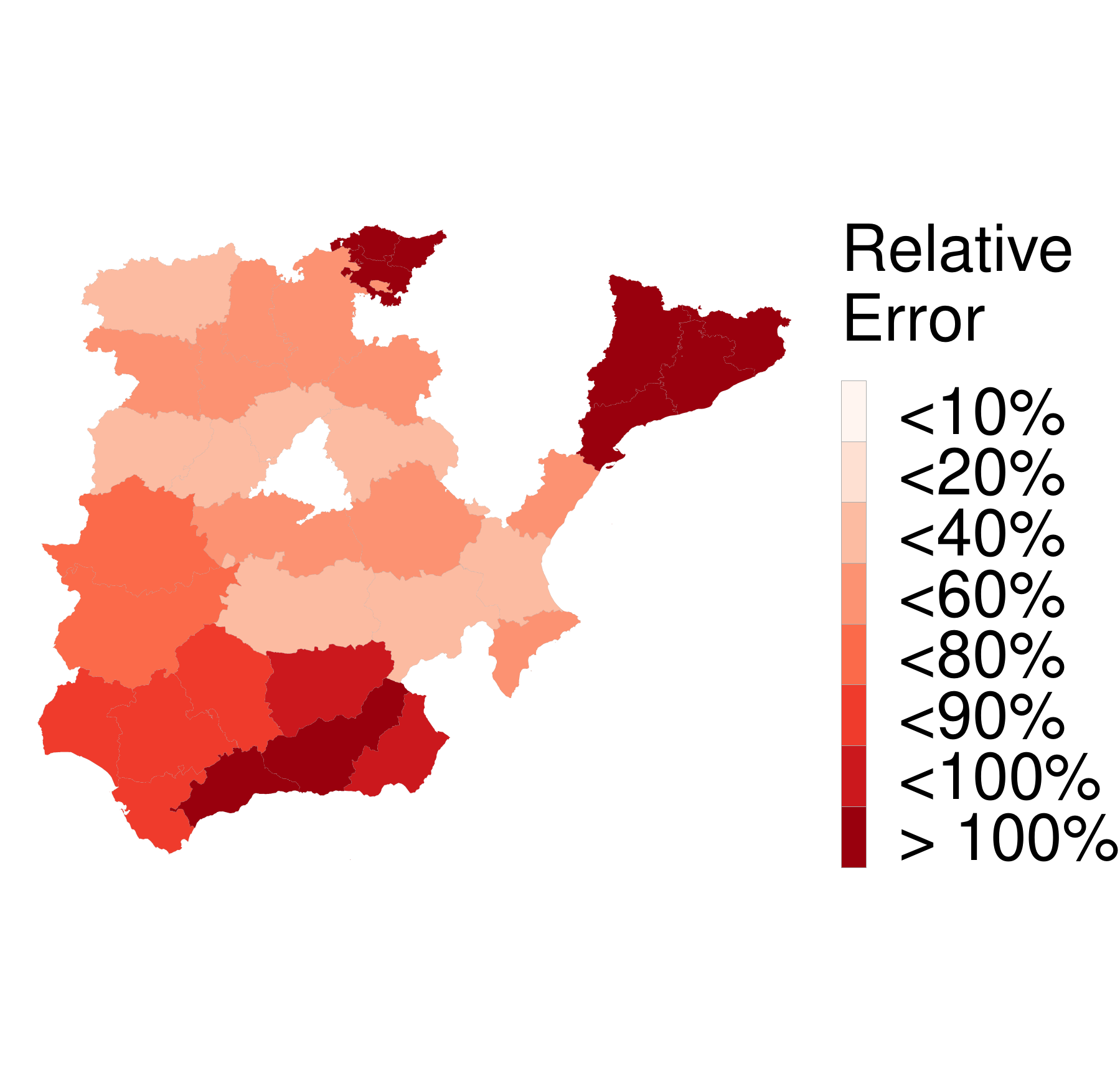}}
\subfigure{
\includegraphics[height=1in]{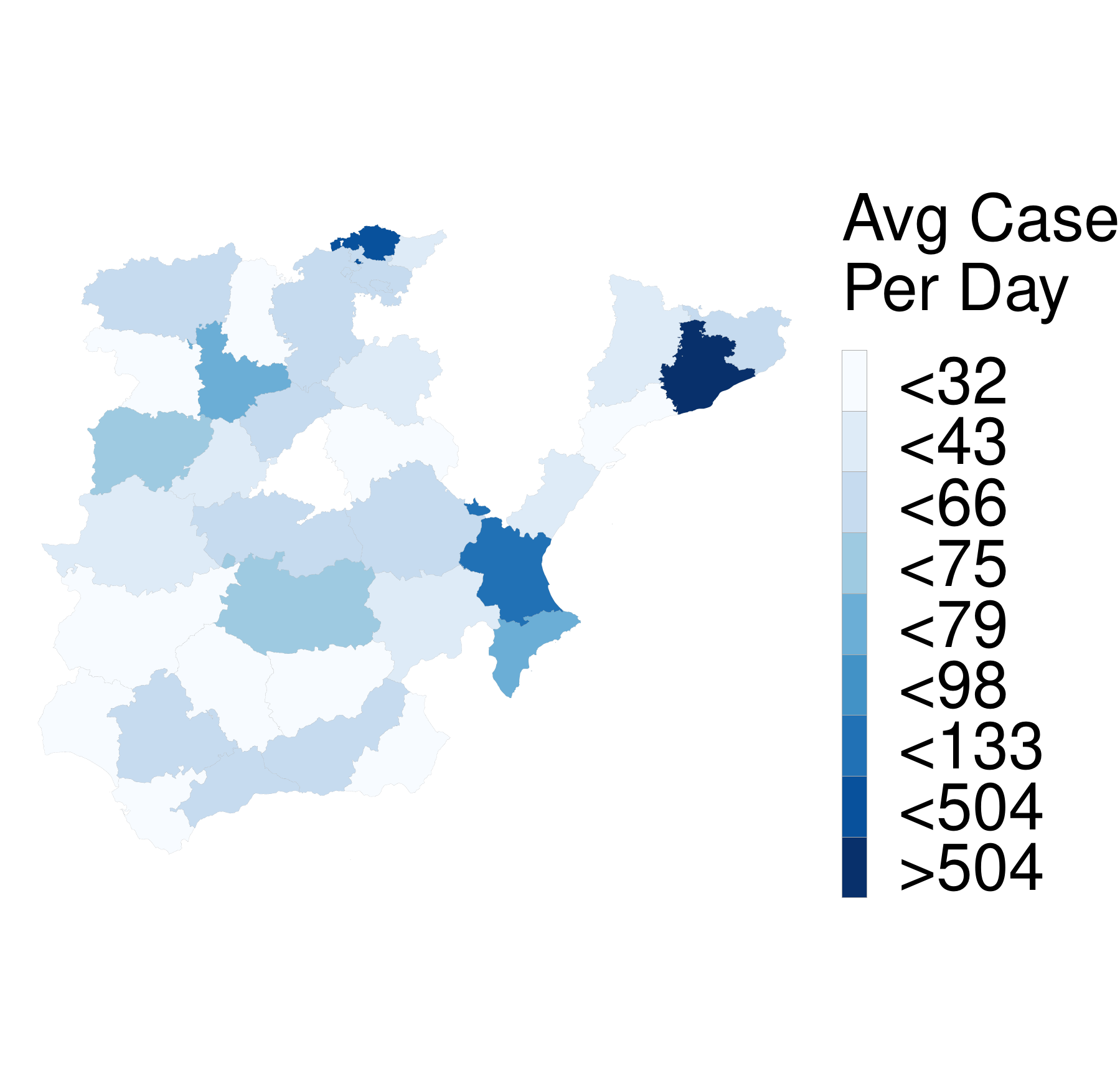}
}
\subfigure{
\includegraphics[height=1.1in]{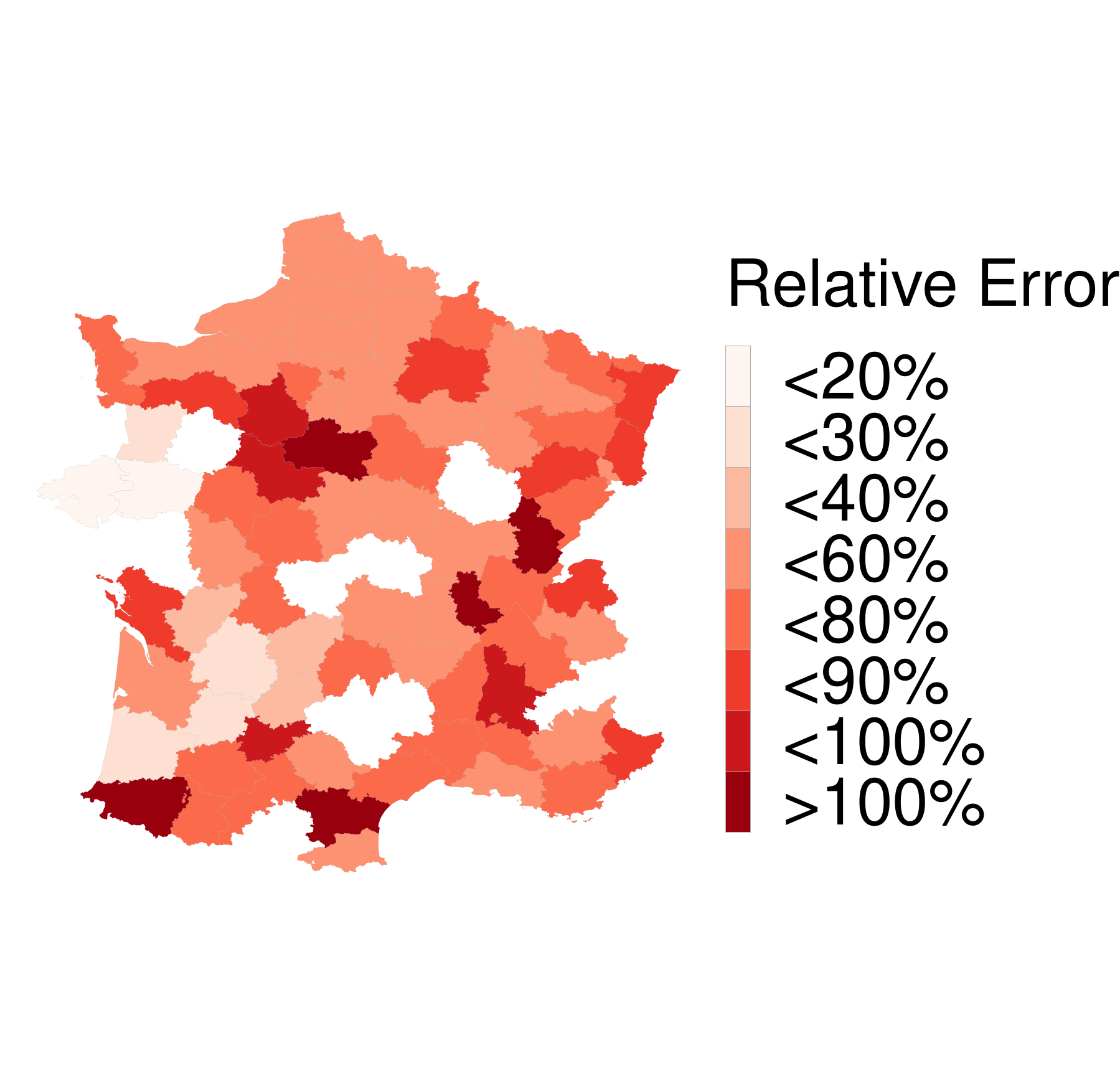}}
%
\subfigure{
\includegraphics[height=1.1in]{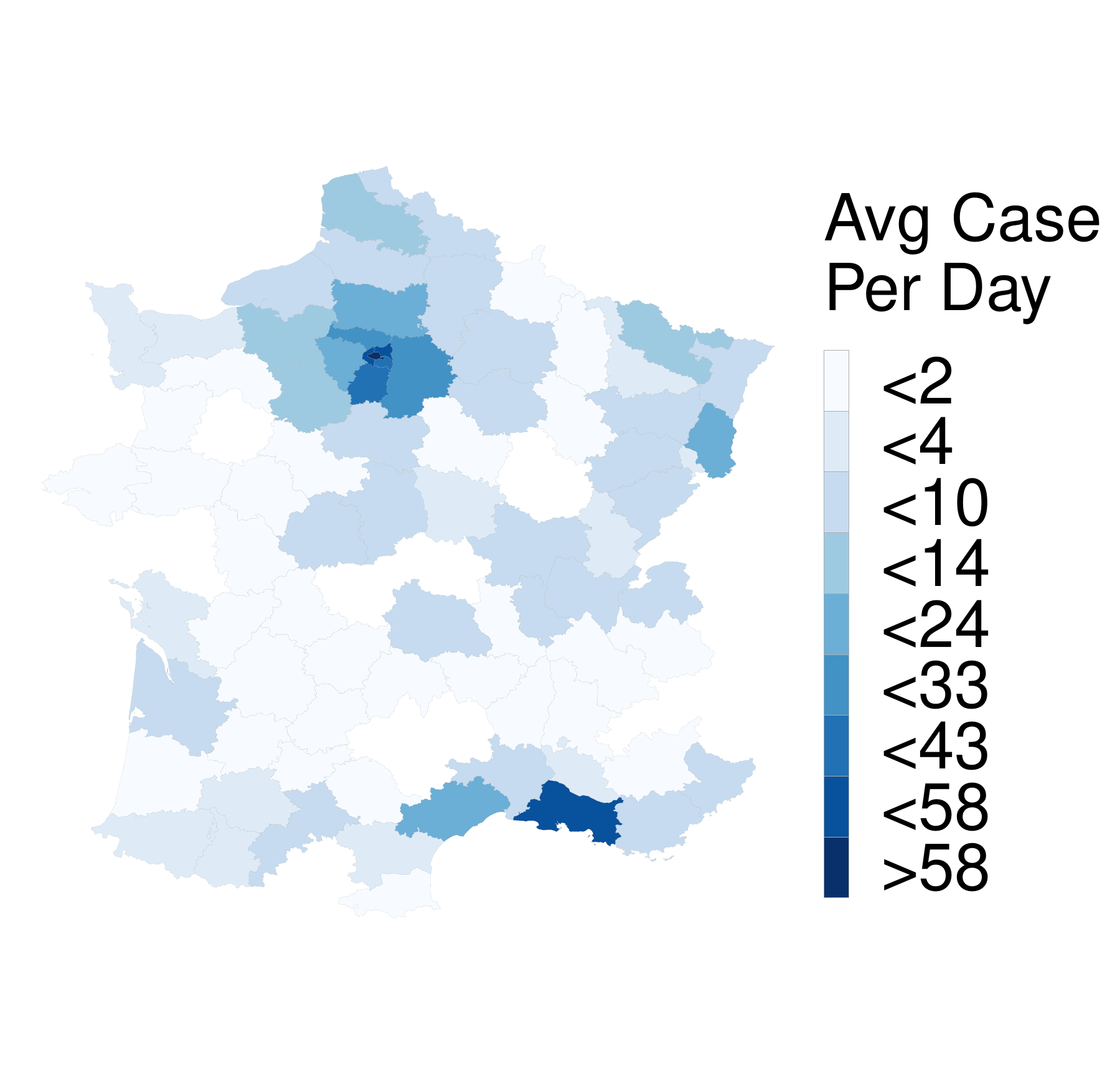}
}
\caption{Plot of the relative test error and average number of cases per day for each available region.}
\label{fig:maps}
\end{figure}

Our error function treats all regions equivalently, independent of the region-specific population and the number of cases, \ie a region with $10$ cases per day should not be treated the same as a region with $1000$.
We need a measure to take into account the region-specific characteristics as well as the time.
Towards this end, we computed the deviation between the average predicted and the actual average number of cases over the next $5$ days (not to be confused with the average error for the next $x$ days in Table~\ref{tab:results}).
The relative error of each region is defined below and is illustrated in Figure~\ref{fig:maps}.
\begin{equation}
   r = \frac{1}{n \, (T-5)}\sum_{t=1}^{T-5} \sum_{v \in V} \frac{\Big| \sum_{i=0}^5 \hat{y}_v^{(t+i)} - \sum_{i=0}^5 y_v^{(t+i)} \Big|}{\sum_{i=0}^5 y_v^{(t+i)}}
\end{equation}
One can see that the regions with high relative error are the ones with the fewest cases. On the other hand, the regions with the highest number of cases tend to have much smaller relative error, less than $20\%$ to be exact, with the exception of one region in Spain.
This indicates that our model indeed produces accurate predictions that could be useful in resource allocation and policy-making during the pandemic.

Fig. \ref{fig:maps} also allows us to evaluate the method more objectively. From a machine learning perspective, one may argue that even though the \textbf{MPNN+TL} outperforms the baselines, their predictions are not very accurate in terms of average error.
This is partially explained due to the inherent problems of the dataset mentioned at sec \ref{sec:dataset} as well as the assumption that case reporting is standard throughout the regions.
We expect a large improvement in performance in case a standard methodology for case reporting is adopted and the number of tests per region remains constant and proportional to the population.
Having said that, utilizing such a model in practice is more than feasible, as the difference in scale is more useful at the regional level. In other words, a region predicted to have 200 new cases total in the next 5 days will have similar needs with a region with 240 or 160 real cases (20\% error). Contrary to that, a prediction of 200 cases with a real value of 300 would be a more significant misclassification, which is not possible looking at the results at Figure~\ref{fig:maps}. Moreover, regions with big relative error tend to have a small number of cases e.g. the model may predict 20 cases in a region with 10 (100\% error), which is also acceptable from a real-world perspective.

\section{Conclusion}\label{sec:conclusion}
In this paper, we presented a model for COVID-19 forecasting which could provide useful insights to policymakers and allow them to make informed decisions on appropriate interventions and resource allocation.
The proposed model builds upon the recent work on GNNs. We use mobility data as a graph where nodes correspond to regions and edge weights to measures of human mobility between their endpoints.
Then, we derive variants of the family of MPNNs to generate representations for the regions based on their interactions.
Furthermore, since different countries might be in different phases of the epidemic, we propose to transfer a well-performing disease spreading model from one country to another where limited data is available.
Experiments conducted on data from $4$ European countries show that our architectures outperform traditional and more recent approaches in predicting the number of daily new COVID-19 cases.

In terms of future directions, we plan to provide our model with \eg demographics regarding the the age/gender distribution and features related to the weather. Furthermore, we will include additional data from Facebook such as the intensity of connectedness between regions measured by the friendship relationships between two regions. 
Our final goal is to evaluate the model on the second wave of COVID-19, based on the first.

\section{Acknowledgements}
This research is supported by i. the French National research agency via the ANR XCOVIF (AAP RA-COVID-19 V6) project  and ii. by Greece and the European Union (European Social Fund- ESF) through the Operational Programme $\ll$ Human Resources Development, Education and Lifelong Learning$\gg$ in the context of the project ``Reinforcement of Postdoctoral Researchers - 2\textsuperscript{nd} Cycle'' (MIS-5033021), implemented by the State Scholarships Foundation (IKY).

\bibliography{bib}

\begin{thebibliography}{27}
\providecommand{\natexlab}[1]{#1}
\providecommand{\url}[1]{\texttt{#1}}
\providecommand{\urlprefix}{URL }
\expandafter\ifx\csname urlstyle\endcsname\relax
  \providecommand{\doi}[1]{doi:\discretionary{}{}{}#1}\else
  \providecommand{\doi}{doi:\discretionary{}{}{}\begingroup
  \urlstyle{rm}\Url}\fi

\bibitem[{Chimmula and Zhang(2020)}]{chimmula2020time}
Chimmula, V. K.~R.; and Zhang, L. 2020.
\newblock Time Series Forecasting of COVID-19 transmission in Canada Using LSTM
  Networks.
\newblock \emph{Chaos, Solitons \& Fractals} 109864.

\bibitem[{Colizza et~al.(2006)Colizza, Barrat, Barthelemy, and
  Vespignani}]{colizza2006prediction}
Colizza, V.; Barrat, A.; Barthelemy, M.; and Vespignani, A. 2006.
\newblock Prediction and predictability of global epidemics: the role of the
  airline transportation network.
\newblock \emph{Bulletin of the American Physical Society} .

\bibitem[{Deng et~al.(2019)Deng, Wang, Rangwala, Wang, and
  Ning}]{deng2019graph}
Deng, S.; Wang, S.; Rangwala, H.; Wang, L.; and Ning, Y. 2019.
\newblock Graph message passing with cross-location attentions for long-term
  ILI prediction.
\newblock \emph{arXiv preprint arXiv:1912.10202} .

\bibitem[{Finn, Abbeel, and Levine(2017)}]{finn2017model}
Finn, C.; Abbeel, P.; and Levine, S. 2017.
\newblock Model-agnostic meta-learning for fast adaptation of deep networks.
\newblock In \emph{34th International Conference on Machine Learning},
  1126--1135.

\bibitem[{Flaxman et~al.(2020)Flaxman, Mishra, Gandy, Unwin, Mellan, Coupland,
  Whittaker, Zhu, Berah, Eaton et~al.}]{flaxman2020estimating}
Flaxman, S.; Mishra, S.; Gandy, A.; Unwin, H. J.~T.; Mellan, T.~A.; Coupland,
  H.; Whittaker, C.; Zhu, H.; Berah, T.; Eaton, J.~W.; et~al. 2020.
\newblock Estimating the effects of non-pharmaceutical interventions on
  COVID-19 in Europe.
\newblock \emph{Nature} 584(7820): 257--261.

\bibitem[{Gao et~al.(2020)Gao, Sharma, Qian, Glass, Spaeder, Romberg, Sun, and
  Xiao}]{gao2020stan}
Gao, J.; Sharma, R.; Qian, C.; Glass, L.~M.; Spaeder, J.; Romberg, J.; Sun, J.;
  and Xiao, C. 2020.
\newblock STAN: Spatio-Temporal Attention Network for Pandemic Prediction Using
  Real World Evidence.
\newblock \emph{arXiv preprint arXiv:2008.04215} .

\bibitem[{Gilmer et~al.(2017)Gilmer, Schoenholz, Riley, Vinyals, and
  Dahl}]{gilmer2017neural}
Gilmer, J.; Schoenholz, S.~S.; Riley, P.~F.; Vinyals, O.; and Dahl, G.~E. 2017.
\newblock Neural message passing for Quantum chemistry.
\newblock In \emph{34th International Conference on Machine Learning},
  1263--1272.

\bibitem[{Graves and Jaitly(2014)}]{graves2014towards}
Graves, A.; and Jaitly, N. 2014.
\newblock {Towards End-to-End Speech Recognition with Recurrent Neural
  Networks}.
\newblock In \emph{31st International Conference on Machine Learning},
  1764--1772.

\bibitem[{Kapoor et~al.(2020)Kapoor, Ben, Liu, Perozzi, Barnes, Blais, and
  O'Banion}]{kapoor2020examining}
Kapoor, A.; Ben, X.; Liu, L.; Perozzi, B.; Barnes, M.; Blais, M.; and O'Banion,
  S. 2020.
\newblock Examining COVID-19 Forecasting using Spatio-Temporal Graph Neural
  Networks.
\newblock In \emph{16th International Workshop on Mining and Learning with
  Graphs}.

\bibitem[{Kipf and Welling(2017)}]{kipf2017semi}
Kipf, T.~N.; and Welling, M. 2017.
\newblock Semi-supervised classification with graph convolutional networks.
\newblock In \emph{5th International Conference on Learning Representations}.

\bibitem[{Kufel(2020)}]{kufel2020arima}
Kufel, T. 2020.
\newblock ARIMA-based forecasting of the dynamics of confirmed Covid-19 cases
  for selected European countries.
\newblock \emph{Equilibrium. Quarterly Journal of Economics and Economic
  Policy} 15(2): 181--204.

\bibitem[{Lampos et~al.(2020)Lampos, Moura, Yom-Tov, Cox, McKendry, and
  Edelstein}]{lampos2020tracking}
Lampos, V.; Moura, S.; Yom-Tov, E.; Cox, I.~J.; McKendry, R.; and Edelstein, M.
  2020.
\newblock Tracking COVID-19 using online search.
\newblock \emph{arXiv preprint arXiv:2003.08086} .

\bibitem[{Lee et~al.(2017)Lee, Kim, Lee, and Yoon}]{lee2017transfer}
Lee, J.; Kim, H.; Lee, J.; and Yoon, S. 2017.
\newblock Transfer learning for deep learning on graph-structured data.
\newblock In \emph{31st AAAI Conference on Artificial Intelligence}.

\bibitem[{Lorch et~al.(2020)Lorch, Trouleau, Tsirtsis, Szanto, Sch{\"o}lkopf,
  and Gomez-Rodriguez}]{lorch2020spatiotemporal}
Lorch, L.; Trouleau, W.; Tsirtsis, S.; Szanto, A.; Sch{\"o}lkopf, B.; and
  Gomez-Rodriguez, M. 2020.
\newblock A spatiotemporal epidemic model to quantify the effects of contact
  tracing, testing, and containment.
\newblock \emph{arXiv preprint arXiv:2004.07641} .

\bibitem[{Maas et~al.(2019)Maas, Iyer, Gros, Park, McGorman, Nayak, and
  Dow}]{maas2019facebook}
Maas, P.; Iyer, S.; Gros, A.; Park, W.; McGorman, L.; Nayak, C.; and Dow, P.~A.
  2019.
\newblock Facebook Disaster Maps: Aggregate Insights for Crisis Response \&
  Recovery.
\newblock In \emph{ISCRAM}.

\bibitem[{Mahmud(2020)}]{mahmud2020bangladesh}
Mahmud, S. 2020.
\newblock Bangladesh COVID-19 Daily Cases Time Series Analysis using Facebook
  Prophet Model.
\newblock \emph{Available at SSRN 3660368} .

\bibitem[{Mallick et~al.(2020)Mallick, Balaprakash, Rask, and
  Macfarlane}]{mallick2020transfer}
Mallick, T.; Balaprakash, P.; Rask, E.; and Macfarlane, J. 2020.
\newblock Transfer Learning with Graph Neural Networks for Short-Term Highway
  Traffic Forecasting.
\newblock \emph{arXiv preprint arXiv:2004.08038} .

\bibitem[{Morris et~al.(2019)Morris, Ritzert, Fey, Hamilton, Lenssen, Rattan,
  and Grohe}]{morris2019weisfeiler}
Morris, C.; Ritzert, M.; Fey, M.; Hamilton, W.~L.; Lenssen, J.~E.; Rattan, G.;
  and Grohe, M. 2019.
\newblock Weisfeiler and leman go neural: Higher-order graph neural networks.
\newblock In \emph{33rd AAAI Conference on Artificial Intelligence},
  4602--4609.

\bibitem[{Nikolentzos, Tixier, and Vazirgiannis(2020)}]{nikolentzos2020message}
Nikolentzos, G.; Tixier, A. J.-P.; and Vazirgiannis, M. 2020.
\newblock Message Passing Attention Networks for Document Understanding.
\newblock In \emph{34th AAAI Conference on Artificial Intelligence},
  8544--8551.

\bibitem[{Paszke et~al.(2019)Paszke, Gross, Massa, Lerer, Bradbury, Chanan,
  Killeen, Lin, Gimelshein, Antiga et~al.}]{paszke2019pytorch}
Paszke, A.; Gross, S.; Massa, F.; Lerer, A.; Bradbury, J.; Chanan, G.; Killeen,
  T.; Lin, Z.; Gimelshein, N.; Antiga, L.; et~al. 2019.
\newblock Pytorch: An imperative style, high-performance deep learning library.
\newblock In \emph{Advances in Neural Information Processing Systems},
  8026--8037.

\bibitem[{Seo et~al.(2018)Seo, Defferrard, Vandergheynst, and
  Bresson}]{seo2018structured}
Seo, Y.; Defferrard, M.; Vandergheynst, P.; and Bresson, X. 2018.
\newblock Structured sequence modeling with graph convolutional recurrent
  networks.
\newblock In \emph{International Conference on Neural Information Processing},
  362--373. Springer.

\bibitem[{Soriano-Panos et~al.(2020)Soriano-Panos, Ghoshal, Arenas, and
  G{\'o}mez-Gardenes}]{soriano2020impact}
Soriano-Panos, D.; Ghoshal, G.; Arenas, A.; and G{\'o}mez-Gardenes, J. 2020.
\newblock Impact of temporal scales and recurrent mobility patterns on the
  unfolding of epidemics.
\newblock \emph{Journal of Statistical Mechanics: Theory and Experiment}
  2020(2): 024006.

\bibitem[{Tang et~al.(2020)Tang, Li, Sun, Yao, Mitra, and
  Wang}]{tang2020transferring}
Tang, X.; Li, Y.; Sun, Y.; Yao, H.; Mitra, P.; and Wang, S. 2020.
\newblock Transferring Robustness for Graph Neural Network Against Poisoning
  Attacks.
\newblock In \emph{13th International Conference on Web Search and Data
  Mining}, 600--608.

\bibitem[{Yao et~al.(2018)Yao, Wu, Ke, Tang, Jia, Lu, Gong, Li, Ye, and
  Chuxing}]{yao2018deep}
Yao, H.; Wu, F.; Ke, J.; Tang, X.; Jia, Y.; Lu, S.; Gong, P.; Li, Z.; Ye, J.;
  and Chuxing, D. 2018.
\newblock Deep multi-view spatial-temporal network for taxi demand prediction.
\newblock In \emph{32nd AAAI Conference on Artificial Intelligence},
  2588--2595.

\bibitem[{Yao et~al.(2019)Yao, Zhang, Wei, Jiang, Wang, Huang, Chawla, and
  Li}]{yao2019graph}
Yao, H.; Zhang, C.; Wei, Y.; Jiang, M.; Wang, S.; Huang, J.; Chawla, N.~V.; and
  Li, Z. 2019.
\newblock Graph few-shot learning via knowledge transfer.
\newblock \emph{arXiv preprint arXiv:1910.03053} .

\bibitem[{Zeroual et~al.(2020)Zeroual, Harrou, Dairi, and
  Sun}]{zeroual2020deep}
Zeroual, A.; Harrou, F.; Dairi, A.; and Sun, Y. 2020.
\newblock Deep Learning Methods for Forecasting COVID-19 Time-Series Data: A
  Comparative Study.
\newblock \emph{Chaos, Solitons \& Fractals} 110121.

\bibitem[{Zou, Lampos, and Cox(2019)}]{zou2019transfer}
Zou, B.; Lampos, V.; and Cox, I. 2019.
\newblock Transfer Learning for Unsupervised Influenza-like Illness Models from
  Online Search Data.
\newblock In \emph{2019 World Wide Web Conference}, 2505--2516.

\end{thebibliography}

\end{document}